\documentclass[journal=nalefd,manuscript=letter]{achemso}

\setkeys{acs}{
	articletitle = true,
	maxauthors = 10
}

\usepackage[T1]{fontenc}
\usepackage[version=3]{mhchem}
\usepackage{graphicx}
\usepackage{amsmath,amssymb}
\usepackage{bm}
\usepackage{booktabs}
\usepackage{multirow}
\usepackage{array}
\usepackage{float}
\usepackage{xcolor}

\title{Strain- and Field-Tunable Nonrelativistic Spin Splitting and Wave-Symmetry–Dependent Spin Transport in Twisted Bilayer Altermagnets}

\author{Shantanu Pathak}
\affiliation{Department of Physics, Indian Institute of Technology Delhi, New Delhi 110016, India}
\email{shantanu.pathak@physics.iitd.ac.in}

\author{Saswata Bhattacharya}
\affiliation{Department of Physics, Indian Institute of Technology Delhi, New Delhi 110016, India}
\email{saswata@physics.iitd.ac.in}

\keywords{Altermagnetism, Twisted bilayers, Nonrelativistic spin splitting, 
	Strain and E-field engineering, Spin transport, $k\cdot p$ model}

\begin{document}

\begin{abstract}
	Magnetism-driven nonrelativistic spin splitting (NRSS) provides a pathway toward efficient, spin–orbit-free spintronics. 
	In centrosymmetric two-dimensional antiferromagnets, spin-polarized transport is symmetry-forbidden due to the combined space–time inversion ($PT$) symmetry. 
	Here, by employing first-principles density functional theory and spin-group symmetry analysis, we demonstrate that twisting two antiferromagnetic or ferromagnetic monolayers of CoCl$_2$, AX$_2$ (A = Mn, V; X = Cl, Br, I), NiF$_2$, NiBr$_2$, FeS, CoS, MnTe$_2$, MnSe$_2$, and RuSe induces finite NRSS even in the absence of spin–orbit coupling. 
	The relative twist breaks $[C_2||P]$ and $[E||C_{nz}]$ symmetries, giving rise to momentum-dependent spin polarization with distinct $d$-, $g$-, and $i$-wave altermagnetic patterns across the Brillouin zone. 
	Using symmetry-invariant $k\cdot p$ modeling, we extract linear spin-splitting coefficients $\alpha^{(1)}$ ranging from 800–1100 meV\AA, comparable to SOC-induced Rashba–Dresselhaus strengths observed in noncentrosymmetric semiconductors. 
	An out-of-plane electric field ($\mathcal{E}_z$) introduces Zeeman-type band splitting up to 110 meV at 10 MV/cm, while biaxial strain tunes the NRSS magnitude nearly linear without altering symmetry. 
	Crucially, the strain $u_{xx-yy}$ reduces the spin point group symmetry and drives reversible $g/i \!\rightarrow\! d$ wave-type transitions, resulting in finite spin conductivity and an enhanced spin-splitter angle (up to 18\textdegree). 
	These results extend the concept of altermagnetism to twisted bilayer geometries and establish a general route for realizing exchange-driven, nonrelativistic spin currents through symmetry engineering—without requiring heavy elements or spin–orbit coupling.
\end{abstract}

	\section{TOC graphic}
\begin{figure}[H] 
	\centering
	\includegraphics[width=0.85\textwidth, height=0.3\textheight]{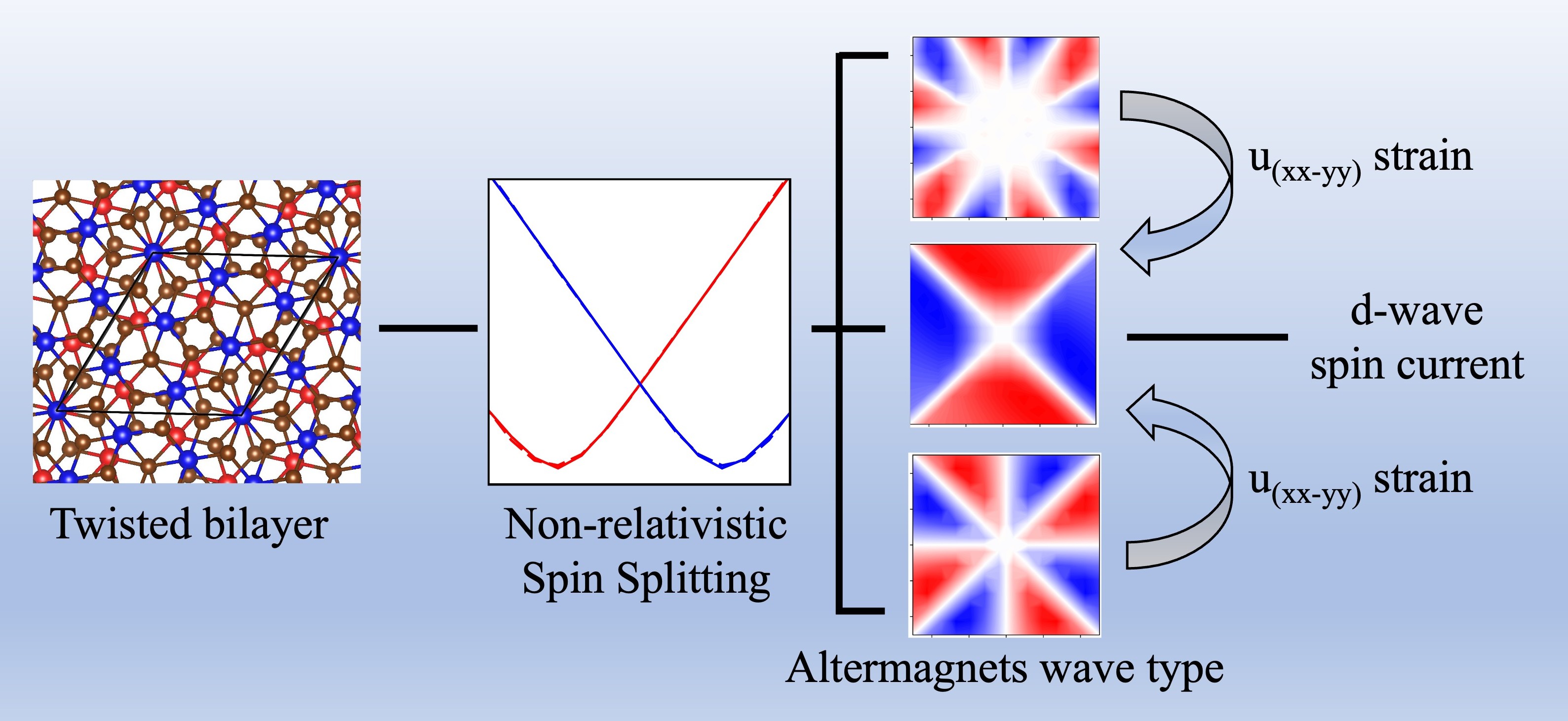}
\end{figure}
	
\maketitle


The generation, manipulation, and detection of pure spin currents constitute a central challenge in modern spintronics~\cite{vzutic2004spintronics, hoffmann2013spin, sinova2015spin}. 
Conventional spintronic devices rely on relativistic spin–orbit coupling (SOC) to achieve spin-to-charge conversion through the spin Hall~\cite{hirsch1999spin, kato2004observation} or Rashba–Edelstein effects~\cite{edelstein1990spin}. 
However, SOC-driven spin transport is inevitably limited by spin relaxation mechanisms such as Elliott–Yafet~\cite{elliott1954theory,yafet1963solid} and D’yakonov–Perel~\cite{dyakonov1972spin}, resulting in short coherence lengths and dissipative spin flow. 
This has motivated an emerging paradigm of \emph{nonrelativistic spintronics}, where spin splitting and spin currents arise purely from exchange interactions rather than relativistic SOC~\cite{sheoran2024nonrelativistic, belashchenko2025giant, smejkal2024altermagnetic, Pathak2025JPCL}.

A key development in this direction is the discovery of \emph{altermagnetism}—a new magnetic phase that bridges the symmetry characteristics of ferromagnets (FMs) and antiferromagnets (AFMs)~\cite{smejkal2022beyond, hayami2020bottom, yuan2021prediction, jungwirth2024altermagnets}. 
Altermagnets exhibit a compensated collinear magnetic order but lack combined space–time inversion ($PT$) symmetry, giving rise to a spin-split band structure entirely in the absence of SOC. 
This \emph{nonrelativistic spin splitting} (NRSS) manifests as momentum-dependent spin polarization with higher-order angular symmetries—typically classified as $d$-, $g$-, or $i$-wave spin textures~\cite{gonzalez2021efficient, vsmejkal2022beyond, sheoran2025tuning, Liu2025NanoLett_2D, Liu2025NanoLett_dwave, Chen2025NanoLett, Ma2025NanoLett_superatoms}. 
While $d$-wave altermagnets such as RuO$_2$ and $\alpha$-Fe$_2$O$_3$ host finite spin-polarized transport~\cite{gonzalez2021efficient, feng2022anomalous}, higher-order $g$- and $i$-wave systems display symmetry-enforced spin degeneracy that suppresses spin conductivity~\cite{belashchenko2025giant, karetta2025strain}. 
Thus, a major theoretical and experimental challenge is to achieve controlled transitions between these altermagnetic states—particularly $g\!\rightarrow\!d$ and $i\!\rightarrow\!d$ symmetry conversions—to unlock dissipationless spin transport~\cite{bose2022tilted, bai2025nature, guo2025hidden}.

External perturbations such as electric fields and lattice strain offer powerful means of tuning altermagnetic order. 
Electric gating can lift $[C_2||P]$ symmetry and induce magnetoelectric-like spin polarization~\cite{che2025bilayer}, while mechanical strain lowers the effective spin point group (SPG), enabling emergent spin conductivity through symmetry reduction~\cite{zhang2025strain, karetta2025strain, ezawa2025hall, Li2026NanoLett_laser}. 
In particular, biaxial strain modifies the magnitude of NRSS without changing symmetry, whereas diagonal strain ($u_{xx-yy}$) can break rotational invariances, driving transitions from $g/i$- to $d$-wave altermagnetic phases~\cite{belashchenko2025giant, karetta2025strain}. 
Such strain-engineered control over NRSS provides a new path toward spin–orbit-free spin current generation~\cite{Zhang2026NanoLett_hidden, Samanta2025NanoLett_filter}.

Two-dimensional (2D) and moiré-twisted bilayer materials are especially promising for realizing tunable altermagnetism due to their symmetry-sensitive stacking geometries and strong interlayer exchange interactions~\cite{liu2024twisted, sheoran2024nonrelativistic, zhou2025moire, qi2024spin}. 
Twisting two AFM or FM layers can naturally remove $PT$ symmetry and generate momentum-asymmetric spin textures at generic $\mathbf{k}$ points, producing NRSS even in the nonrelativistic limit~\cite{sheoran2024nonrelativistic, liu2024twisted}. 
This “twistronic altermagnetism” opens new routes to engineer exchange-driven spin splitting, tunable by twist angle, electric field, or strain, without relying on heavy elements or SOC.

In this work, we perform a systematic first-principles and symmetry-based investigation of twisted bilayers composed of CoCl$_2$, AX$_2$ (A = Mn, V; X = Cl, Br, I), NiF$_2$, NiBr$_2$, FeS, CoS, MnTe$_2$, MnSe$_2$, and RuSe. 
We show that these systems host robust NRSS with characteristic $d$-, $g$-, and $i$-wave patterns across the Brillouin zone. 
By applying out-of-plane electric fields and various strain configurations, we demonstrate controllable tuning of the linear spin-splitting coefficient and reversible $g/i \!\rightarrow\! d$ wave-type transitions. 
The resulting modulation of spin conductivity and spin-splitter angle establishes twisted bilayer altermagnets as a unified platform for exploring symmetry-engineered, exchange-driven spin transport—offering new design principles for spin–orbit-free spintronics and 2D straintronics.

\section{Symmetry Analysis}

Before proceeding to a detailed symmetry analysis, 
we first provide a global classification of two-dimensional magnetic materials 
based on their magnetic order, magnetic space group (MSG), 
and the presence or absence of nonrelativistic spin splitting (NRSS), 
as summarized in Table~\ref{tab:NRSS_classification}. 
The classification highlights the spin-group symmetries that protect spin degeneracy 
or allow its lifting at generic $\mathbf{k}$ points, 
in both monolayer and twisted bilayer geometries.

Of particular relevance to this work are ferromagnetic systems (SST-2), 
which exhibit NRSS already at the monolayer level, 
and antiferromagnetic systems with MSG type~III (SST-3), 
where spin degeneracy is symmetry-protected in the monolayer 
but is generically lifted by twisting. 
These two classes encompass all materials investigated here 
and provide a natural starting point for the symmetry analysis below, 
which focuses on collinear compensated magnets 
and the symmetry-lowering mechanisms induced by strain.

\begin{table*}[t]
	\caption{Classification of two-dimensional materials based on magnetic space group (MSG) type, magnetic order, and their impact on nonrelativistic spin splitting (NRSS). 
		The spin-group symmetry protecting spin degeneracy at generic $\mathbf{k}$ points is indicated in parentheses. 
		Only materials explicitly investigated in this work are listed as examples unless otherwise noted.}
	\label{tab:NRSS_classification}
	\centering
	
	\resizebox{\textwidth}{!}{
		\begin{tabular}{cccccc}
			\toprule
			Spin-splitting & Magnetic & MSG & \multicolumn{2}{c}{NRSS at generic $\mathbf{k}$} & \multirow{2}{*}{Examples} \\
			\cmidrule(lr){4-5}
			type & order & type & Monolayer & Twisted bilayer &  \\
			\midrule
			SST-1 & Nonmagnetic & II 
			& $\times \,([\mathcal{C}_{2}\Vert E])$ 
			& $\times \,([\mathcal{C}_{2}\Vert E])$ 
			& MoS$_2$~\cite{Xiao2012MoS2}, PtSe$_2$~\cite{Zhang2017PtSe2} \\
			
			SST-2 & Ferromagnetic & I / III 
			& $\checkmark$ 
			& $\checkmark$ 
			& CoCl$_2$ (Both phases),  \\
			
			&  &  
			&  
			&  
			& AX$_2$ (A = Mn, V; X = Cl, Br, I), \\
			
			&  &  
			&  
			&  
			& NiF$_2$, NiBr$_2$, MnTe$_2$, MnSe$_2$ (this work) \\
			
			SST-3 & Antiferromagnetic & III 
			& $\times \,([\mathcal{C}_{2}\Vert \mathcal{P}]/[\mathcal{C}_{2}\Vert \mathcal{M}_z])$ 
			& $\checkmark$ 
			& FeS, CoS, RuSe (this work) \\
			
			SST-4 & Altermagnetic & I / III 
			& $\checkmark$ 
			& -- 
			& -- \\
			
			SST-5 & Antiferromagnetic & IV 
			& $\times \,([\mathcal{C}_{2}\Vert \boldsymbol{\tau}])$ 
			& -- 
			& -- \\
			\bottomrule
		\end{tabular}
	}
\end{table*}

Collinear compensated magnets can be systematically classified using spin-group symmetry operations of the form $[R_1||R_2]$, where $R_1$ and $R_2$ act on the spin and real-space coordinates, respectively~\cite{litvin1974spin,vsmejkal2022beyond}. 
This formalism allows an explicit treatment of magnetic symmetry in the nonrelativistic limit, where spin and real spaces are decoupled. 
The spin transformation under a combined operation is expressed as
\begin{equation}
	\mathbf{D}(R_2)\, \sigma^{\mathbf{D}(R_1)s_k} \, \mathbf{D}(R_2)^{\dagger} = \sigma^{s_k},
\end{equation}
where $\sigma^{s_k}$ represents the spin current operator along the spin-polarization direction $s_k$. 
A particular component $\sigma^{s_k}_{ij}$ of the spin-conductivity tensor is symmetry-allowed only when the spin point group (SPG) does not contain any operation that enforces $\sigma^{s_k}_{ij} = \sigma^{-s_k}_{ij}$.

In antiferromagnets (AFMs), opposite spin sublattices are related by the $[C_2||P]$ symmetry, which maps $(\mathbf{k}, s) \rightarrow (-\mathbf{k}, -s)$ and restores Kramers-like spin degeneracy in the absence of spin–orbit coupling, thereby forbidding a spin current. 
In contrast, altermagnets (AMs) are characterized by mirror-rotation symmetries, such as $[C_2||C_{2x}]$, $[C_2||C_{2y}]$, or $[C_2||C_{4z}]$, that connect opposite spin sublattices. 
Depending on the rotational order, these correspond to $d$-, $g$-, or $i$-wave altermagnetic orders~\cite{chen2024_spingroup_prb110}. 
The presence or absence of these operations determines whether spin currents are symmetry-allowed.

Strain fields $\eta_{ij}$ transform as even-parity, second-rank tensors ($k_i k_j$) under inversion and rotation operations. 
While biaxial strain ($\eta_{xx+yy} = \eta_{xx}$ and $\eta_{yy}$) preserves the rotational and mirror symmetries of the parent SPG, shear ($\eta_{xy}$) or diagonal ($\eta_{xx-yy} =\eta_{xx}$ and $\eta_{-yy}$) strains can reduce these symmetries, leading to transitions between altermagnetic classes. 
The reduction of the parent SPG to a lower-symmetry subgroup enables finite spin conductivity by breaking the operations that previously connected opposite spin sublattices.

In the planar $g$-wave altermagnet with the parent SPG $^14^22^22$, the system possesses rotational symmetries $[E||C_{2x}]$, $[E||C_{2y}]$, and $[E||C_{4z}]$, which connect the opposite spin sublattices and suppress spin-polarized transport. 
The application of diagonal strain $\eta_{xx-yy}$ or shear strain $\eta_{xy}$ breaks the $[E||C_{4z}]$ symmetry, thereby reducing the group to $^12^22^22$, where the remaining rotational operations are $[E||C_{2x}]$ and $[E||C_{2z}]$. 
This reduced configuration corresponds to a $d$-wave altermagnet that allows finite spin conductivity $\sigma^s_{xy}$.

\begin{table*}[t]
	\caption{Tuning altermagnetism by the application of strain. The first and second columns denote the altermagnetic pattern and the parent SPG, respectively. The latter six columns denote the effective SPG of the parent space group after the application of the strain field $\eta_{ij}$. Transitions labeled $(d)$ represent cases where the symmetry reduces to a $d$-wave configuration, allowing finite spin conductivity.}
	\centering
	\renewcommand{\arraystretch}{1.2}
	
	\resizebox{\textwidth}{!}{
		\begin{tabular}{c c c c c c c c}
			\toprule
			Altermagnetic Pattern	& \textbf{SPGs} & $\eta_{xx}$ & $\eta_{yy}$ & $\eta_{zz}$ & $\eta_{xy}$ & $\eta_{xz}$ & $\eta_{yz}$ \\
			\midrule
			Planar $g$-wave	&	$^14/^1m^2m^2m$  & $^2m^2m^1m$ $(d)$ & $^2m^2m^1m$ $(d)$ & $^14/^1m^2m^2m$ $(g)$  & $^2m^2m^1m$ $(d)$ & $^22/^2m$ $(d)$ & $^22/^2m$ $(d)$ \\
			Planar $g$-wave	   & $^14^2m^2m$	& $^2m^2m^12$ $(d)$ & $^2m^2m^12$ $(d)$ & $^14^2m^2m$ $(g)$ & $^2m^2m^12$ $(d)$ & $^2m$ $(d)$ & $^2m$ $(d)$\\
			Planar $g$-wave	   & $^14^22^22$ & $^22^22^12$ $(d)$ & $^22^22^12$ $(d)$ & $^14^22^22$ $(g)$ & $^22^22^12$ $(d)$ & $^22$ $(d)$ & $^22$ $(d)$ \\
			Planar $g$-wave	   & $^1\overline{4} {^2}2^2m$ & $^22^22^12$ $(d)$ &  $^22^22^12$ $(d)$& $^1\overline{4} {^2}2^2m$ $(g)$ & $^2m^2m^12$ $(d)$& $^22$ $(d)$ & $^22$ $(d)$\\
			Planar $i$-wave		  &  $^16/^1m^2m^2m$ & $^2m^2m^1m$ $(d)$ & $^2m^2m^1m$ $(d)$ & $^16/^1m^2m^2m$ $(i)$ & $^12/^1m$ & $^22/^2m$ $(d)$ & $^22/^2m$ $(d)$ \\
			Planar $i$-wave		 &   $^16^2m^2m$ & $^2m^2m^12$ $(d)$ & $^2m^2m^12$ $(d)$ & $^16^2m^2m$ $(i)$ & $^12$ $(d)$ & $^2m$ $(d)$ & $^2m$ $(d)$\\
			Planar $i$-wave		 &   $^16^22^22$& $^22^22^12$ $(d)$ & $^22^22^12$ $(d)$ & $^16^22^22$ $(i)$ & $^12$ $(d)$ & $^22$ $(d)$ & $^22$ $(d)$\\
			Planar $i$-wave		 &   $^1\overline{6}{^2}m^22$& $^2m^22^1m$ $(d)$ & $^2m^22^1m$ $(d)$ & $^1\overline{6}{^2}m^22$ $(i)$ & $^1m$ & $^22$ $(d)$ & $^2m$ $(d)$ \\
			\bottomrule
		\end{tabular}
	}
	
	\label{tab:strain_symmetry}
\end{table*}

A similar strain-driven mechanism occurs in planar $i$-wave systems described by SPG $^16^22^22$, where the same symmetry-breaking strain components convert the system to an effective $^12^22^22$ ($d$-wave) symmetry. 
In both cases, biaxial strain preserves the original $g$- or $i$-wave symmetry (tuning only the magnitude of spin splitting), while diagonal or shear strain lowers the symmetry class, activating finite spin conductivity.

In summary, the strain-driven reduction of the spin point group provides a microscopic route for tuning altermagnetic order in two-dimensional systems. 
While biaxial strain acts as a linear tuning parameter for the spin-splitting magnitude, diagonal and shear strains play a crucial role in inducing transitions from $g/i$-wave to $d$-wave altermagnets, thereby enabling the emergence of nonrelativistic spin currents in twisted bilayer systems.

To further understand the microscopic origin of the nonrelativistic spin splitting (NRSS) revealed by the symmetry analysis, we examine the Brillouin zone (BZ) structure and construct a symmetry-invariant $k\cdot p$ Hamiltonian based on the little-group operations at the band-splitting points.

For the hexagonal twisted bilayers, such as tb-MnTe$_2$ and tb-CoCl$_2$, the reduced moiré Brillouin zone preserves the conventional $\Gamma$–M–K–$\Gamma$ path, where spin-degenerate bands are observed. 
To capture the NRSS observed along the generic directions, we define additional symmetry-equivalent paths $M_1$–$M_c$–$M_2$ and $K_1$–$K_c$–$K_2$, where $M_c$ and $K_c$ are the central points exhibiting maximal spin splitting. 
The little-group symmetry of the $M_c$ and $K_c$ points consists of $[C_2||C_{2[010]}]$. 
This rotational symmetry, combined with the exchange field of the collinear compensated order, determine the allowed invariants in the $k\cdot p$ Hamiltonian~\cite{bir1974symmetry,dresselhaus1955spin}.

Similarly, for orthorhombic-type twisted bilayers (e.g., tb-CoCl$_2$, tb-FeS, tb-RuSe, tb-CoS), the reciprocal-space path follows $\Gamma$–X–M–Y–$\Gamma$, where the $M$ point represents the corner of the square Brillouin zone, analogous to the $M$ point in the hexagonal case. 
To analyze the NRSS observed along the X–Y direction, we define the midpoint $M_c$ along this path as the symmetry-related splitting point, consistent with the notation used throughout this work. 
The symmetry elements of the $M_c$ point belong to the same $[C_2||C_{2[010]}]$ class as those of $M_c$ in the hexagonal lattice; therefore, the invariant Hamiltonian form remains identical for both lattice families.

\begin{figure}[t]
	\includegraphics[width=0.9\columnwidth]{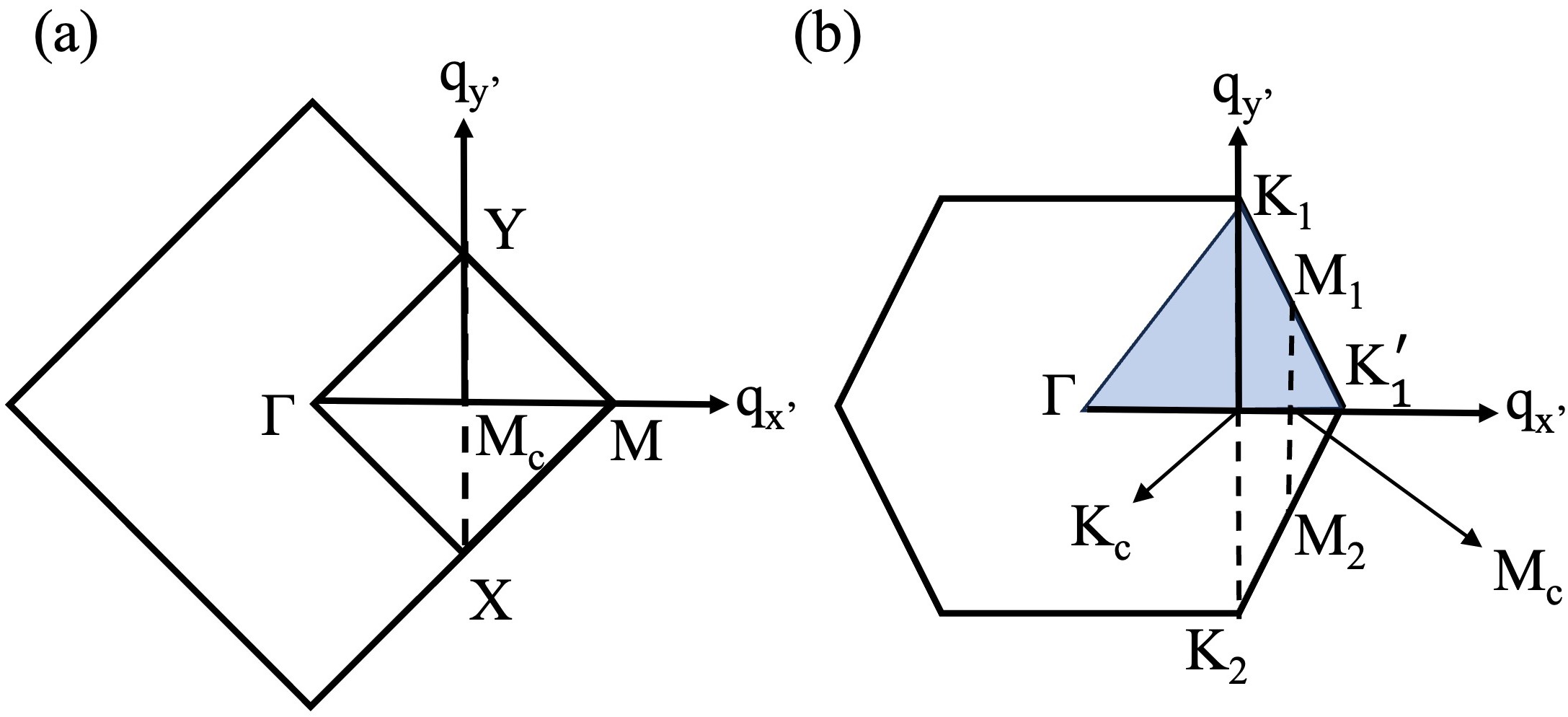}
	\caption{
		Definition of the local momentum coordinates used in the symmetry-based $k\cdot p$ analysis of twisted bilayers.
		(a) Square (tetragonal) moir\'e Brillouin zone, showing the generic point $M_c$ along the $X$–$M$–$Y$ direction.
		The local coordinate system $(q_x',q_y')$ is defined with respect to $M_c$.
		(b) Hexagonal moir\'e Brillouin zone, highlighting the generic point $K_c$ along the $K_1$–$K_2$ direction and the generic point $M_c$ along the $M_1$–$M_2$ direction,
		with $q_x'$ and $q_y'$ defined relative to $K_c$ and $M_c$.
		The shaded regions indicate the irreducible momentum sectors used to construct the symmetry-allowed
		nonrelativistic $k\cdot p$ Hamiltonian.
	}
	\label{fig:kp_bz}
\end{figure}

In the local coordinate system $(x',y')$ defined around the generic symmetry-reduced points
$M_c$ or $K_c$ in the moir\'e Brillouin zone
(see Fig.~\ref{fig:kp_bz}),
the momentum deviation $\mathbf{q}=\mathbf{k}-\mathbf{k}_c$ provides a natural basis
for constructing the lowest-order symmetry-invariant terms.
For both hexagonal and tetragonal twisted bilayers, the residual symmetry at these points
is governed by a two-fold rotation combined with a real-space rotation,
denoted as $[C_2||C_{2[010]}]$.
Under this operation, the momentum components transform anisotropically,
which forbids spin splitting along $q_{x'}$ while allowing odd-order terms in $q_{y'}$.

Taking these symmetry constraints into account, the effective nonrelativistic
$k\cdot p$ Hamiltonian near $M_c$ or $K_c$ can be written as
\begin{equation}
	H_{\mathrm{eff}} =
	\frac{\hbar^2 q_{x'}^2}{2m_{x'}} +
	\frac{\hbar^2 q_{y'}^2}{2m_{y'}} +
	\left( \alpha^{(1)} q_{y'} + \alpha^{(3)} q_{y'}^3 \right)\sigma_z ,
	\label{eq:heff_full}
\end{equation}
where $m_{x'}$ and $m_{y'}$ are the effective carrier masses along the principal axes,
and $\alpha^{(1)}$ and $\alpha^{(3)}$ denote the linear and cubic
spin-splitting coefficients, respectively.
The first two terms describe the parabolic band dispersion near the NRSS point,
while the odd powers of $q_{y'}$ encode the spin-dependent splitting
permitted by the reduced spin-point-group symmetry.

The linear term in $q_{y'}$ produces a clear energy separation between spin-up
and spin-down bands, analogous in form to a Rashba--Dresselhaus splitting,
but originating purely from exchange-driven symmetry breaking rather than
spin--orbit coupling.
The cubic contribution $\alpha^{(3)}q_{y'}^3$ introduces higher-order anisotropy
away from the degeneracy point, which becomes relevant at larger momentum deviations.

To quantitatively extract the spin-splitting parameters,
we fit the DFT-calculated energy eigenvalues $E_i(\mathbf{q})$
to the analytical eigenvalues of $H_{\mathrm{eff}}$ using a weighted
least-squares procedure,
\begin{equation}
	S=\sum_{i=1}^{2}\sum_{\mathbf{q}} f(\mathbf{q})
	\left|
	\det\!\left[
	H_{\mathrm{eff}}(\mathbf{q})-E_i(\mathbf{q})I
	\right]
	\right|^2 ,
\end{equation}
where the weight function $f(\mathbf{q})$ follows a Gaussian distribution
centered at $\mathbf{q}=0$.
This choice emphasizes the fitting accuracy near the spin-degenerate point
while suppressing overfitting at large $|\mathbf{q}|$.

The invariant Hamiltonian in Eq.~(\ref{eq:heff_full}) thus provides a unified
microscopic description of nonrelativistic spin splitting in twisted bilayer
altermagnets across both hexagonal (MnTe$_2$, CoCl$_2$) and tetragonal
(FeS, CoS, RuSe, CoCl$_2$) lattice families.
In all cases, the same $[C_2||C_{2[010]}]$ symmetry governs the anisotropy of the
spin splitting, confirming that the observed NRSS is an intrinsic consequence
of altermagnetic exchange symmetry rather than relativistic spin--orbit effects.

\section{Results and Discussion}

\subsection{Electronic structure and nonrelativistic spin splitting}

We constructed twisted bilayers (tb-) from fully optimized monolayers of CoCl$_2$, AX$_2$ (A = Mn, V, Ni; X = Cl, Br, I), FeS, CoS, MnTe$_2$, MnSe$_2$, and RuSe. 
These compounds crystallize in either trigonal or tetragonal symmetries, resulting in distinct lattice geometries and spin-group symmetries in the two-dimensional limit. 
The trigonal systems derive from the space group $P\bar{3}m1$ (No.~162), while the tetragonal systems originate from $P4/mmm$ (No.~123) or $I4/mmm$ (No.~139). 
All calculations were performed in the nonrelativistic limit, without spin–orbit coupling, such that spin and real-space symmetry operations remain decoupled within the spin-group formalism~\cite{vsmejkal2022beyond,sheoran2024nonrelativistic}.

\begin{figure*}[t]
	\includegraphics[width=\textwidth, height=0.4\textheight]{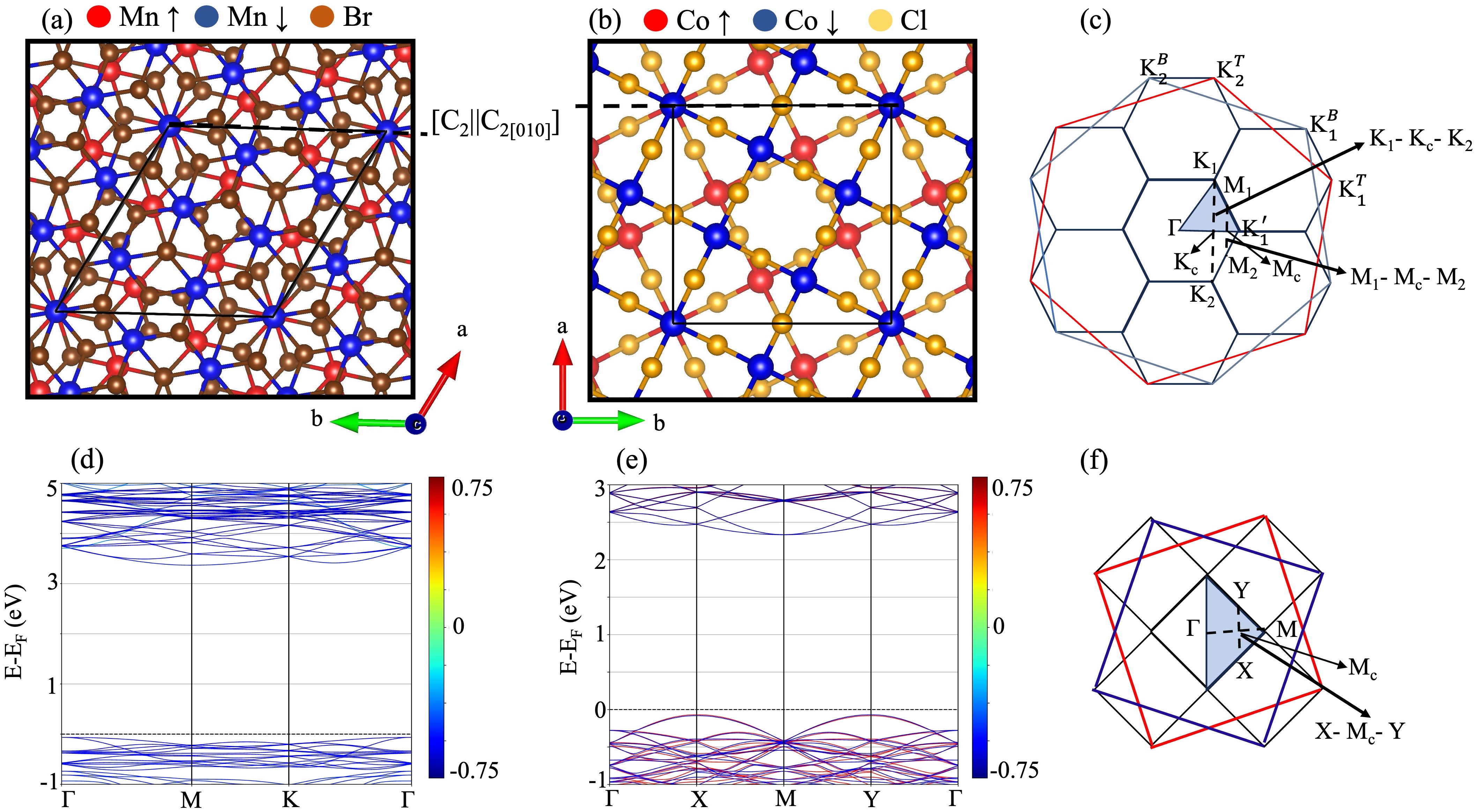}
	\caption{
		Crystal structures, moir\'e Brillouin zones (BZs), and spin-resolved band structures of representative twisted bilayer altermagnets.
		(a) Relaxed moir\'e superlattice of hexagonal tb-MnBr$_2$ and (b) tetragonal tb-CoCl$_2$, where red and blue spheres denote opposite spin-polarized transition-metal atoms and halogen atoms are shown in brown (Br) and yellow (Cl).
		Dashed lines indicate the nonrelativistic spin-group symmetry operation $[C_2||C_{2[010]}]$ connecting opposite spin sublattices.
		(c,f) Construction of the moir\'e BZ from the individual layer BZs (red/blue hexagons or squares for hexagonal/tetragonal lattices) to the reduced moir\'e BZ (black), with high-symmetry paths indicated.
		(d,e) Spin-resolved band structures of tb-MnBr$_2$ along $\Gamma$–M–K–$\Gamma$ and tb-CoCl$_2$ along $\Gamma$–X–M–Y–$\Gamma$, calculated within PBE.
		Although spin degeneracy persists along high-symmetry lines, the reduced symmetry of twisted bilayers enables nonrelativistic spin splitting at generic $\mathbf{k}$ points.
	}
	\label{fig:structure_bz_bands}
\end{figure*}

Each monolayer was structurally relaxed until the residual forces on all atoms were below $10^{-3}$~eV/\AA. 
The magnetic ground state of each system was determined by comparing total energies for several collinear configurations, including ferromagnetic, layer-antiferromagnetic, and Néel-type antiferromagnetic orders. 
Although several monolayers exhibit intralayer ferromagnetic alignment, the energetically preferred bilayer configuration is interlayer antiferromagnetic in all cases, resulting in an overall compensated magnetic state. 
The calculated local magnetic moments are 4.613~$\mu_B$ (Mn), 2.694~$\mu_B$ (Co), 2.731~$\mu_B$ (V), 1.623~$\mu_B$ (Ni), 2.702~$\mu_B$ (Ru), and 3.502~$\mu_B$ (Fe), consistent with high-spin $d$-electron configurations reported for related transition-metal compounds.

Commensurate twisted bilayers were generated starting from the relaxed AA-stacked bilayer as the untwisted reference configuration. 
For trigonal (hexagonal) lattices, the commensurate twist angles $\theta$ satisfy the coincidence lattice condition~\cite{moon2014commensurate}
\begin{equation}
	\cos \theta = \frac{n^2 + 4mn + m^2}{2(m^2 + mn + n^2)},
	\label{eq:twist_trigonal}
\end{equation}
where $m$ and $n$ are integers defining the moiré supercell. 
A twist angle of $\theta = 21.78^{\circ}$ yields a stable commensurate structure with 42 atoms per supercell and is adopted throughout this work for trigonal systems. 
For tetragonal lattices, commensurate twisting is achieved through a square-lattice rotation described by~\cite{can2021geometry}
\begin{equation}
	\cos \theta = \frac{m^2 - n^2}{m^2 + n^2},
	\label{eq:twist_tetragonal}
\end{equation}
which produces a commensurate structure at $\theta = 53.13^{\circ}$ for $(m,n)=(3,4)$. 
These twist angles substantially reduce the crystalline symmetry, break the combined $\mathcal{P}\mathcal{T}$ invariance of the untwisted bilayers, and generate nontrivial moiré patterns in both lattice families.

We examined several possible interlayer spin configurations, including $\uparrow\uparrow\uparrow\uparrow$, $\uparrow\downarrow\uparrow\downarrow$, $\uparrow\downarrow\downarrow\uparrow$, and $\uparrow\uparrow\downarrow\downarrow$, where the arrows denote the relative spin polarization on each magnetic sublattice. 
Although the magnetic ground state of the individual monolayers may be either ferromagnetic or antiferromagnetic depending on the compound, the lowest-energy twisted bilayer configuration corresponds to a collinearly compensated state with antiferromagnetic interlayer coupling. 
For monolayers with intrinsic antiferromagnetic order, this corresponds to the $\uparrow\downarrow\uparrow\downarrow$ arrangement featuring both intralayer and interlayer compensation, whereas for ferromagnetic monolayers it corresponds to an $\uparrow\uparrow\downarrow\downarrow$ stacking of oppositely polarized layers. 
In all cases, the resulting bilayer exhibits zero net magnetization, in which opposite spin sublattices are connected by combined spin–lattice symmetry operations of the form $[\mathcal{C}_2||\mathcal{C}_{2i}]$ ($i=x,y,z$), depending on the lattice geometry. 
As a result, the twisted bilayers realize altermagnetic order characterized by nontrivial spin-group symmetries in the nonrelativistic limit.

Representative relaxed moir\'e superlattice structures of twisted bilayer MnBr$_2$ and CoCl$_2$ are shown in
Fig.~\ref{fig:structure_bz_bands}(a,b).
In both cases, twisting reduces the crystalline symmetry while preserving collinear magnetic order,
resulting in compensated spin textures connected by nonrelativistic spin-group operations such as
$[C_2||C_{2[010]}]$.
The corresponding moir\'e Brillouin zones are constructed from the Brillouin zones of the individual layers
[Fig.~\ref{fig:structure_bz_bands}(c,f)],
giving rise to new high-symmetry points ($K_c$, $M_c$) and generic momentum paths that are absent in untwisted bilayers.
Along conventional high-symmetry paths—$\Gamma$–M–K–$\Gamma$ for trigonal lattices and
$\Gamma$–X–M–Y–$\Gamma$ for tetragonal lattices—the spin-resolved band structures
[Fig.~\ref{fig:structure_bz_bands}(d,e)]
remain doubly degenerate, consistent with symmetry protection by operations such as
$[\mathcal{C}_2||\mathcal{P}]$ and $[\mathcal{C}_2||\tau]$.
In contrast, along generic momentum paths including
K$_1$–K$_\mathrm{c}$–K$_2$ and M$_1$–M$_\mathrm{c}$–M$_2$ for trigonal systems,
and X–Y for tetragonal systems, a clear energy splitting between spin-up and spin-down states emerges.
This nonrelativistic spin splitting originates purely from exchange interactions and the symmetry reduction induced by twisting,
without any contribution from spin–orbit coupling, and forms the basis for the quantitative analysis presented below.

The magnitude and anisotropy of the NRSS depend sensitively on the lattice symmetry and magnetic configuration, with tetragonal systems generally exhibiting stronger splitting due to their lower rotational symmetry. 
The absence of spin splitting along high-symmetry directions and its emergence only along generic $\mathbf{k}$-paths confirm the symmetry-protected nature of NRSS, in full agreement with predictions from nonrelativistic spin-group theory~\cite{sheoran2024nonrelativistic,liu2024twisted}. 
In the following sections, we quantify the NRSS using symmetry-derived $k\cdot p$ models and analyze its tunability under external electric fields and strain.

\subsection{Nonrelativistic Spin Splitting, Momentum-Space Symmetry, and Altermagnetic Classification}

To quantify the nonrelativistic spin splitting (NRSS) and correlate it with the underlying momentum-space symmetry of twisted bilayer altermagnets, 
we performed symmetry-constrained $k\cdot p$ fittings of the DFT-calculated band structures near the characteristic splitting points 
($K_c/M_c$ for hexagonal lattices and $M_c$ for tetragonal lattices). 
A comprehensive quantitative summary of the extracted parameters for all studied systems is provided in Table~\ref{tab:nrss_coeff}. 
The table reports the linear and cubic spin-splitting coefficients, $\alpha^{(1)}$ and $\alpha^{(3)}$, 
for both the valence-band maximum (VBM) and conduction-band minimum (CBM), 
together with the corresponding altermagnetic wave symmetry ($d$-, $g$-, or $i$-wave) 
identified from the momentum-space spin texture and nodal structure.

\begin{figure*}[t]
	\includegraphics[width=\textwidth]{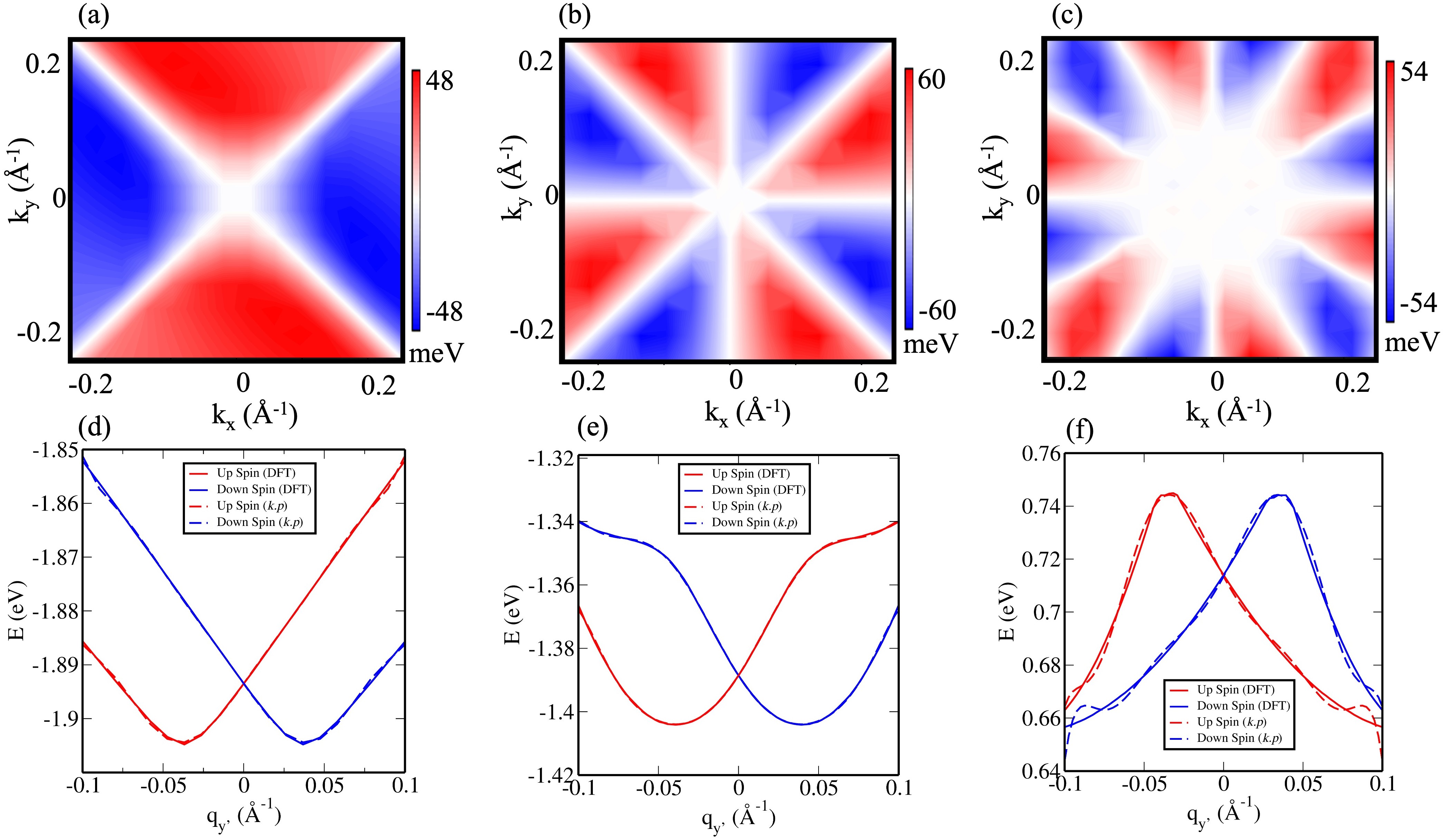}
	\caption{
		Nonrelativistic spin splitting (NRSS) and symmetry-resolved $k\cdot p$ analysis in representative twisted bilayer altermagnets.
		(a) Momentum-resolved spin-splitting energy $\Delta E(\mathbf{k})$ at the VBM of tetragonal tb-FeS, exhibiting a $d$-wave altermagnetic pattern.
		(b) $\Delta E(\mathbf{k})$ at the CBm of tetragonal tb-CoCl$_2$, showing a $g$-wave symmetry.
		(c) $\Delta E(\mathbf{k})$ at the CBm of hexagonal tb-MnBr$_2$, characteristic of an $i$-wave altermagnetic texture.
		(d) Spin-resolved band dispersion near $M_c$ for tb-FeS (VBM), comparing DFT results (solid lines) with the symmetry-invariant $k\cdot p$ model (dashed lines).
		(e,f) Same comparison for the CBm of tetragonal tb-CoCl$_2$ and hexagonal tb-MnBr$_2$, respectively.
		The excellent agreement validates the symmetry-derived effective Hamiltonian and confirms the nonrelativistic origin of the observed spin splitting.
	}
	\label{fig:nrss_kp}
\end{figure*}

\begin{table*}[t]
	\caption{
		Linear ($\alpha^{(1)}$) and cubic ($\alpha^{(3)}$) spin--splitting coefficients
		extracted from symmetry--resolved band dispersions.
		For each entry, values are reported as VBM / CBM.
		The \emph{Path} column specifies the high--symmetry momentum line along which the
		spin splitting is evaluated:
		$K_1$--$K_c$--$K_2$ and $M_1$--$M_c$--$M_2$ for hexagonal lattices, and
		$X$--$M_c$--$Y$ for tetragonal lattices.
		The AM wave type denotes the symmetry of the altermagnetic spin texture.
	}
	\label{tab:spin_split}
	\centering
	\small
	\begin{tabular}{l c c c c c}
		\toprule
		Compound &
		Lattice &
		AM wave type &
		Path &
		$\alpha^{(1)}$ (meV\,\AA) &
		$\alpha^{(3)}$ (eV\,\AA$^3$) \\
		
		\midrule
		MnTe$_2$ & Hexagonal & i-wave &
		$K_1$--$K_c$--$K_2$ &
		99.87 / 697.16 &
		84.67 / 9.47 \\
		
		&  &  &
		$M_1$--$M_c$--$M_2$ &
		85.35 / 228.24 &
		17.78 / 4.19 \\
		
		MnSe$_2$ & Hexagonal & i-wave &
		$K_1$--$K_c$--$K_2$ &
		45.43 / 626.32 &
		1.69 / 87.17 \\
		
		&  &  &
		$M_1$--$M_c$--$M_2$ &
		107.71 / 387.30 &
		0.18 / 15.05 \\
		
		MnBr$_2$ & Hexagonal & i-wave &
		$K_1$--$K_c$--$K_2$ &
		487.69 / 112.18 &
		56.41 / 263.08 \\
		
		&  &  &
		$M_1$--$M_c$--$M_2$ &
		220.16 / 1120.57 &
		9.87 / 21.11 \\
		
		MnI$_2$ & Hexagonal & i-wave &
		$K_1$--$K_c$--$K_2$ &
		693.28 / 124.61 &
		89.61 / 264.03 \\
		
		&  &  &
		$M_1$--$M_c$--$M_2$ &
		585.18 / 1238.17 &
		52.36 / 22.45 \\
		
		MnCl$_2$ & Hexagonal & i-wave &
		$K_1$--$K_c$--$K_2$ &
		320.19 / 444.94 &
		17.71 / 12.13 \\
		
		&  &  &
		$M_1$--$M_c$--$M_2$ &
		127.07 / 242.09 &
		4.74 / 0.72 \\
		
		VI$_2$ & Hexagonal & i-wave &
		$K_1$--$K_c$--$K_2$ &
		54.12 / 524.19 &
		11.01 / 136.21 \\
		
		&  &  &
		$M_1$--$M_c$--$M_2$ &
		152.71 / 36.72 &
		16.51 / 0.71 \\
		
		VCl$_2$ & Hexagonal & i-wave &
		$K_1$--$K_c$--$K_2$ &
		80.26 / 423.16 &
		13.91 / 68.72 \\
		
		&  &  &
		$M_1$--$M_c$--$M_2$ &
		40.62 / 150.02 &
		0.37 / 15.52 \\
		
		NiF$_2$ & Hexagonal & i-wave &
		$K_1$--$K_c$--$K_2$ &
		240.02 / 71.99 &
		8.58 / 13.33 \\
		
		&  &  &
		$M_1$--$M_c$--$M_2$ &
		48.87 / 38.21 &
		2.38 / 7.64 \\
		
		NiBr$_2$ & Hexagonal & i-wave &
		$K_1$--$K_c$--$K_2$ &
		384.19 / 263.20 &
		34.91 / 18.12 \\
		
		&  &  &
		$M_1$--$M_c$--$M_2$ &
		445.98 / 99.63 &
		12.74 / 0.26 \\
		
		CoCl$_2$ & Hexagonal & i-wave &
		$K_1$--$K_c$--$K_2$ &
		371.09 / 18.63 &
		44.61 / 1.41 \\
		
		&  &  &
		$M_1$--$M_c$--$M_2$ &
		217.24 / 53.11 &
		4.90 / 5.72 \\
		
		CoCl$_2$ & Tetragonal & g-wave &
		$X$–$M_c$–$Y$ &
		659.00 / 787.20 &
		25.10 / 12.32 \\
		
		CoS & Tetragonal & d-wave &
		$X$–$M_c$–$Y$ &
		52.81 / 685.12 &
		1.59 / 31.04 \\
		
		FeS & Tetragonal & d-wave &
		$X$–$M_c$–$Y$ &
		422.05 / 259.14 &
		53.20 / 15.10 \\
		
		RuSe & Tetragonal & d-wave &
		$X$–$M_c$–$Y$ &
		115.12 / 35.42 &
		4.91 / 1.54 \\
		\bottomrule
	\end{tabular}
	\label{tab:nrss_coeff}
\end{table*}

The fitting procedure employed the effective Hamiltonian [Eq.~(\ref{eq:heff_full})], 
which captures both the parabolic band dispersion and the symmetry-allowed linear and cubic NRSS terms. 
Finite values of $\alpha^{(1)}$ and/or $\alpha^{(3)}$ are obtained for all systems, 
confirming the emergence of NRSS in the absence of spin--orbit coupling. 
Among the hexagonal systems, tb-CoCl$_2$ and tb-MnTe$_2$ exhibit particularly large linear coefficients $\alpha^{(1)}$, 
reflecting strong exchange-driven spin polarization along the $K_1$--$K_c$--$K_2$ and $M_1$--$M_c$--$M_2$ directions. 
In these cases, the comparatively smaller cubic contribution $\alpha^{(3)}$ indicates an almost symmetric spin splitting around the NRSS point.

By contrast, systems belonging to the higher-order $i$-wave class, such as tb-VI$_2$ and several Mn- and Ni-based halides, 
display enhanced cubic coefficients $\alpha^{(3)}$, consistent with the presence of nodal lines or surfaces 
protected by rotational or mirror symmetries~\cite{vsmejkal2022beyond}. 
In tetragonal $d$-wave systems such as tb-FeS, tb-CoS, and tb-RuSe, 
both linear and cubic terms coexist, but the reduced symmetry allows finite spin polarization 
and spin conductivity already at linear order.

The symmetry-invariant $k\cdot p$ Hamiltonian reproduces the DFT dispersions near $M_c$ with high fidelity
[Fig.~\ref{fig:nrss_kp}(d)--(f)], confirming that the extracted $\alpha^{(1)}$ and $\alpha^{(3)}$
coefficients capture the essential NRSS physics. The extracted linear spin-splitting coefficients span a wide range, 
from several tens of meV$\cdot$\AA{} up to several hundred meV$\cdot$\AA{}, 
depending on the compound, lattice symmetry, and band edge. 
These values are comparable to, and in many cases exceed, 
experimentally reported Rashba-type spin splittings in relativistic systems, 
such as KTaO$_3$ (10 meV$\cdot$\AA{})~\cite{nakamura2012kto}, 
LaAlO$_3$/SrTiO$_3$ interfaces (4.3 meV$\cdot$\AA{})~\cite{caviglia2010laosto}, 
InGaAs/InAlAs quantum wells (70 meV$\cdot$\AA{})~\cite{koralek2009ingaas}, 
and monolayer MoSSe (77 meV$\cdot$\AA{})~\cite{zhang2014mosse}. 
Unlike Rashba or Dresselhaus effects, which originate from relativistic spin--orbit interaction, 
the NRSS observed here is purely nonrelativistic and arises from the exchange field 
and symmetry lowering inherent to altermagnetic order~\cite{gonzalez2021efficient}.

Representative momentum-resolved spin-splitting maps for $d$-, $g$-, and $i$-wave twisted bilayers
are shown in Fig.~\ref{fig:nrss_kp}(a)--(c),
demonstrating the distinct nodal structures imposed by spin-group symmetry. The fitted energy differences $\Delta E(\mathbf{k})$ range from a few meV near the $M_c$ points 
to several tens of meV across the Brillouin zone, 
depending on the material and local twist-induced symmetry breaking. 
These results establish a clear quantitative link between momentum-space symmetry, 
altermagnetic wave type, and NRSS strength in twisted bilayer systems, 
and provide a realistic basis for experimental detection via angle-resolved photoemission spectroscopy (ARPES) 
and spin-polarized transport measurements.

\subsection{Electric-field tuning Zeeman-type spin splitting}

Controlling spin splitting through an external electric field provides a powerful mechanism to manipulate 
spin-polarized electronic states in two-dimensional (2D) materials, especially in twisted bilayers 
where symmetry can be locally broken by gating~\cite{chang2014electric, li2016electrical, wang2024electric}. 
To explore the effect of electrostatic gating on the nonrelativistic spin splitting (NRSS), 
we performed self-consistent DFT simulations of the twisted bilayer MnTe$_2$ (tb-MnTe$_2$) 
under a uniform out-of-plane electric field ($\mathcal{E}_z$), implemented using the dipole-layer correction 
method introduced by Neugebauer and Scheffler~\cite{neugebauer1992adsorbate}. 

In the absence of spin–orbit coupling, the applied field interacts with the layer-resolved charge distribution, 
breaking inversion symmetry between the upper and lower Te–Mn–Te planes. 
This imbalance induces a spin-dependent energy shift that can be modeled by a Zeeman-like term in the Hamiltonian:
\begin{equation}
	\hat{H}_Z = \lambda\,\mathcal{E}_z\,\sigma_z,
\end{equation}
where $\lambda$ is the coupling constant describing the field–spin interaction strength. 
The resulting spin-resolved eigenvalues,
\begin{equation}
	E_{\pm} = E_0 \pm \lambda\,\mathcal{E}_z,
\end{equation}
describe two oppositely polarized subbands whose energy separation increases linearly with $\mathcal{E}_z$. 
This is analogous to a Zeeman effect but originates entirely from nonrelativistic exchange-field imbalance between the layers, 
rather than from spin–orbit coupling.

\begin{figure*}[t]
	\includegraphics[width=\textwidth]{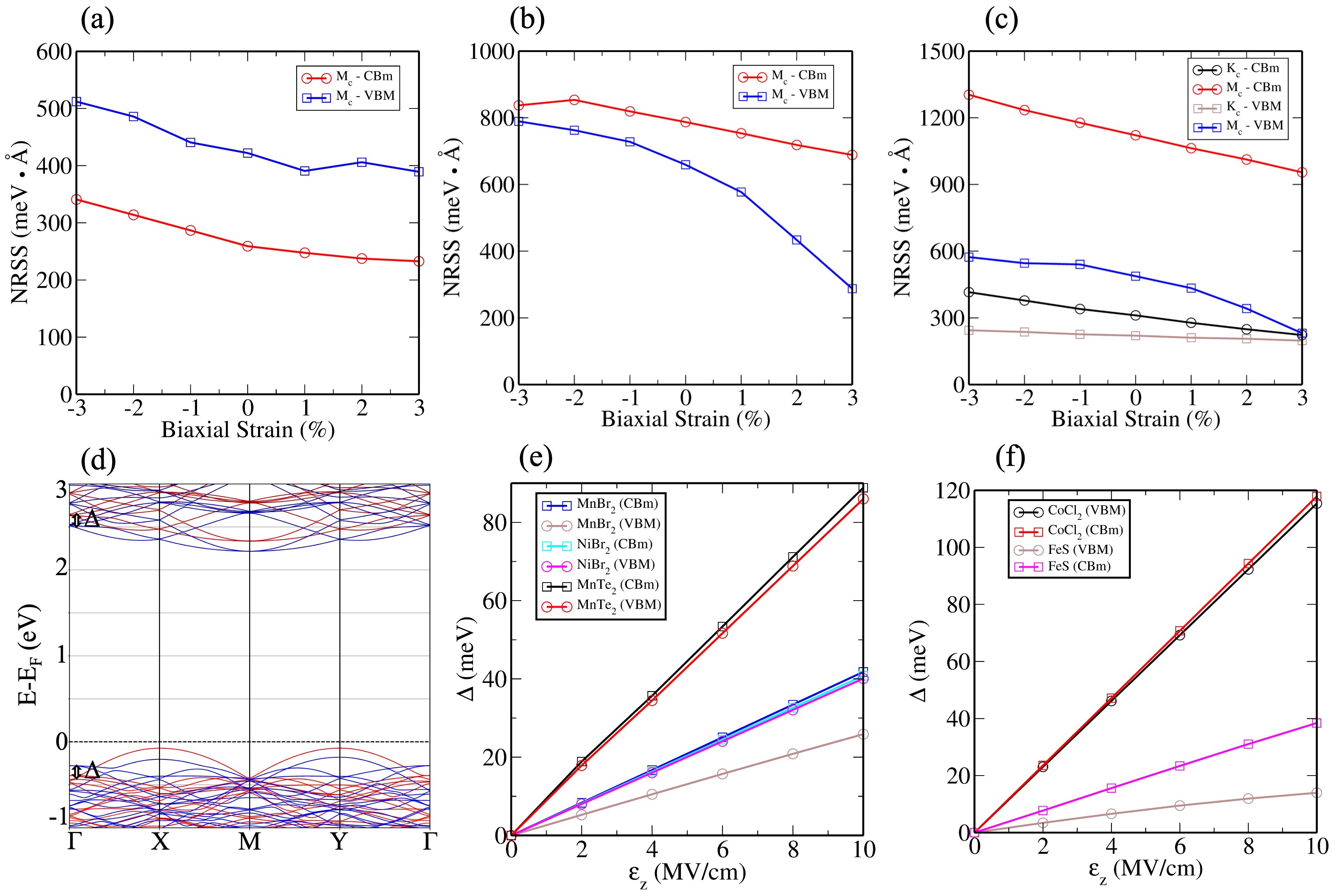}
	\caption{
		Tuning of nonrelativistic spin splitting by biaxial strain and electric field.
		(a--c) Linear spin-splitting coefficient $\alpha^{(1)}$ as a function of biaxial strain
		for tetragonal tb-FeS, tetragonal tb-CoCl$_2$, and hexagonal tb-MnBr$_2$, respectively.
		Both compressive and tensile strain preserve crystal symmetry but continuously tune the NRSS magnitude.
		(d) Spin-resolved band structure of tetragonal tb-FeS under an out-of-plane electric field
		$\mathcal{E}_z = 10$ MV/cm, showing Zeeman-type spin splitting at the $\Gamma$ point.
		(e,f) Electric-field-induced spin splitting $\Delta$ at the VBM and CBm for representative
		hexagonal and tetragonal twisted bilayers, demonstrating a nearly linear dependence on $\mathcal{E}_z$, repectively.
	}
	\label{fig:strain_efield}
\end{figure*}

The calculated band structures of tb-MnTe$_2$ under increasing $\mathcal{E}_z$ (0–10 MV/cm) 
reveal a clear field-induced spin splitting at the $\Gamma$ point. 
At $\mathcal{E}_z = 10$ MV/cm, the splitting magnitude reaches approximately $85$~meV at the valence band maximum (VBM) 
and $90$~meV at the conduction band minimum (CBM), as shown in Fig.~\ref{fig:strain_efield}(e). An out-of-plane electric field induces a Zeeman-type spin splitting at high-symmetry points
[Fig.~\ref{fig:strain_efield}(d)], with the splitting magnitude increasing approximately linearly
with field strength [Fig.~\ref{fig:strain_efield}(e,f)].

The slightly higher splitting at the CBm arises from the dominant in-plane $p$- and $d$-orbital contributions 
that exhibit greater charge asymmetry across the bilayer, making them more sensitive to the out-of-plane field. 
In contrast, VBM states with more out-of-plane character partially screen the electric potential, 
resulting in a smaller splitting amplitude.

This electric-field-induced Zeeman-type splitting in tb-MnTe$_2$ demonstrates that nonrelativistic collinear magnets 
can host large, controllable spin splitting purely through electrostatic gating. 
The linear relationship between $\Delta E_Z = E_\uparrow - E_\downarrow$ and $\mathcal{E}_z$ confirms the field-induced 
exchange imbalance between opposite spin sublattices, with an estimated coupling coefficient 
$\lambda \approx 9$~meV$\cdot$cm/MV from the slope of the fitted data. 
Importantly, this mechanism does not rely on spin–orbit interaction and therefore avoids spin relaxation 
and dephasing effects that typically limit coherence in relativistic spin-split systems.

The magnitude of the obtained Zeeman-type NRSS in tb-MnTe$_2$ (up to 90 meV at 10 MV/cm) and in tb-CoCl$_2$ (tetragonal phase, up to 117 meV at 10 MV/cm) is comparable to, and in some cases exceeds, the Rashba-induced spin-splitting energies reported in heavy-element 2D systems such as monolayer MoSSe and BiTeI~\cite{zhang2014mosse,ishizaka2011bitei}. The nearly linear electric-field dependence of the spin splitting for other systems is shown in Fig.~S3(a--i). These results demonstrate that electric-field control in twisted bilayer altermagnets provides a robust and purely nonrelativistic pathway for generating and tuning spin-polarized transport states, paving the way toward magnetoelectric and low-dissipation spintronic applications.

\subsection{Biaxial strain dependence of NRSS}

In addition to electric-field control, the strength of nonrelativistic spin splitting (NRSS) 
can be tuned through biaxial in-plane strain in twisted bilayer altermagnets. 
We applied uniform biaxial strain ranging from $-3\%$ (compressive) to $+3\%$ (tensile) 
to all systems while preserving their crystal symmetry. 
Because the in-plane strain does not reduce symmetry, 
no additional splitting appears in the high-symmetry paths; 
however, the magnitude of the linear spin-splitting coefficient $\alpha^{(1)}$ 
around the $K_c/M_c$ points varies significantly with strain. 

In all studied systems, $\alpha^{(1)}$ increases under compressive strain 
and decreases under tensile strain, showing nearly linear behavior within the applied range. The linear spin-splitting coefficient $\alpha^{(1)}$ exhibits a monotonic dependence on biaxial strain
for both tetragonal and hexagonal twisted bilayers, as shown in Fig.~\ref{fig:strain_efield}(a--c), with additional representative systems presented in Fig.~S2(a--k).
This trend arises from enhanced interlayer hybridization and exchange interaction 
under lattice compression, and their weakening under tensile strain. 
These results confirm that biaxial strain provides an efficient and reversible means 
to modulate NRSS in twisted bilayers without altering the underlying altermagnetic symmetry.

\subsection{Diagonal strain ($u_{xx-yy}$) and wave-type transition in twisted bilayers}

The application of anisotropic in-plane strain provides a direct route to manipulate 
the symmetry of altermagnetic order in twisted bilayer systems. 
Unlike biaxial strain, which preserves rotational invariance, 
the diagonal strain component $u_{xx-yy}$ breaks the $C_4$ or $C_6$ rotational symmetry 
that protects higher-order ($g$- or $i$-wave) altermagnetic states, 
leading to a reduction in the magnetic spin point group and an associated 
transition toward a $d$-wave symmetry~\cite{karetta2025strain, smejkal2025tuning}. 

In our tetragonal tb-CoCl$_2$, the parent spin point group $^14^22^22$ 
(possessing $[E||C_{2x}]$, $[E||C_{2y}]$, and $[E||C_{4z}]$ rotational symmetries) 
is reduced to $^22^22^12$ upon application of $u_{xx-yy}$ strain. 
This reduction breaks the four-fold rotational symmetry, 
eliminating operations that connect opposite spin-sublattices, 
and induces a transition from a planar $g$-wave to a $d$-wave altermagnetic configuration. 
The $k_x$–$k_y$ resolved spin-splitting energy difference, 
$\Delta E(\mathbf{k}) = E_\uparrow(\mathbf{k}) - E_\downarrow(\mathbf{k})$, 
shows a distinct two-fold symmetry consistent with the $d$-wave character. 

Similarly, in the hexagonal tb-MnTe$_2$, the parent group $^26^22^22$ 
is reduced to $^12^22^22$ under the same $u_{xx-yy}$ strain, 
transforming the planar $i$-wave pattern into a $d$-wave symmetry. 
The contour plots of $\Delta E(\mathbf{k})$ in the $k_x$–$k_y$ plane 
clearly reveal the strain-driven suppression of six-fold spin-splitting lobes 
and the emergence of a two-fold anisotropy characteristic of $d$-wave altermagnetism. 

\begin{figure*}[ht]
	\includegraphics[width=\textwidth]{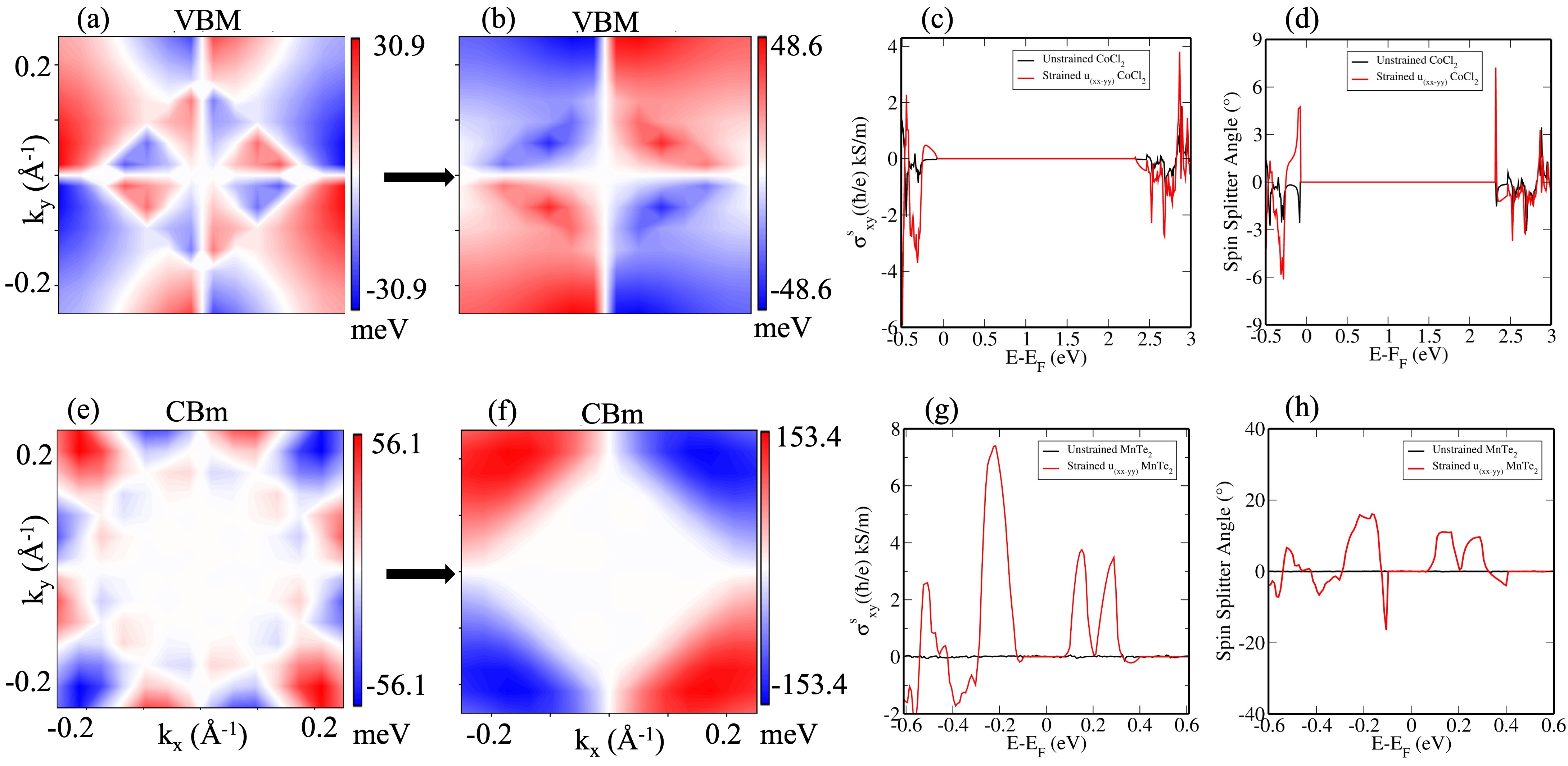}
	\caption{
		Uniaxial strain-induced wave-type transition and emergence of spin-polarized transport.
		(a) $g$-wave momentum-space spin-splitting texture at the VBM of tetragonal tb-CoCl$_2$.
		(b) Corresponding texture under $u_{xx-yy}=1\%$ strain, showing a symmetry-driven $g \rightarrow d$ transition.
		(c,d) Spin conductivity $\sigma^z_{xy}$ and spin-splitter angle (SSA) as a function of energy
		for unstrained and strained tb-CoCl$_2$.
		(e) $i$-wave spin-splitting texture at the CBm of hexagonal tb-MnTe$_2$.
		(f) Conversion to a $d$-wave pattern under $u_{xx-yy}=1\%$ strain.
		(g,h) Corresponding spin conductivity and SSA for tb-MnTe$_2$.
		Uniaxial strain activates finite spin currents and enhances charge-to-spin conversion
		without invoking spin--orbit coupling.
	}
	\label{fig:wave_transition}
\end{figure*}

Furthermore, we find that the magnitude of the spin-splitting energy difference 
$\Delta E(\mathbf{k})$ increases significantly under the applied strain. 
For both tb-CoCl$_2$ and tb-MnTe$_2$, $\Delta E(\mathbf{k})$ approximately doubles 
at the critical $u_{xx-yy}=1\%$ strain compared to the unstrained case, 
indicating that uniaxial strain not only alters the symmetry but also strengthens the effective exchange field 
responsible for NRSS, thereby amplifying spin polarization across the Brillouin zone. The application of diagonal strain $u_{xx-yy}$ reduces the rotational symmetry of the twisted bilayers,
driving a transition from higher-order $g$- or $i$-wave textures to a $d$-wave configuration
[Fig.~\ref{fig:wave_transition}(a,b,e,f)]. 
This strain-controlled $g/i \rightarrow d$ wave transition thus reflects 
a symmetry-governed evolution of nonrelativistic spin splitting, 
in line with recent theoretical predictions for CrSb and related systems~\cite{karetta2025strain}. 
In both tb-CoCl$_2$ and tb-MnTe$_2$, the transition is fully reversible with strain 
and occurs without invoking spin–orbit coupling, 
demonstrating that purely mechanical symmetry control can effectively reconfigure 
the momentum-space spin texture and enhance spin splitting in twisted bilayer altermagnets.

\subsection{Spin and charge conductivity under $u_{xx-yy}$ strain}

As discussed in the previous section, the uniaxial strain component $u_{xx-yy}$ lowers the symmetry of the twisted bilayers,
leading to the $g \rightarrow d$ and $i \rightarrow d$ wave-type transitions. 
To investigate the corresponding transport response, we computed the spin and charge conductivities
($\sigma^{z}_{ij}$ and $\sigma_{ij}$) within the constant relaxation time approximation, using the formalism described in Computational Methods.
The calculations were performed for both the unstrained and strained ($u_{xx-yy} = 1\%$) configurations
of tb-CoCl$_2$ (planar $g \rightarrow d$) and tb-MnTe$_2$ (planar $i \rightarrow d$).

In the unstrained systems, the spin conductivity $\sigma^{z}_{ij}$ vanishes due to the presence of rotational symmetries
($C_4$ in tetragonal and $C_6$ in hexagonal structures) that connect opposite spin-sublattices.
Upon application of $u_{xx-yy}$ strain, these symmetries are broken,
allowing finite spin conductivity components consistent with the reduced $d$-wave spin point groups
($^22^22^12$ and $^12^22^22$ for tb-CoCl$_2$ and tb-MnTe$_2$, respectively).
The charge conductivity, on the other hand, remains nearly unchanged,
demonstrating that the symmetry reduction primarily activates spin-polarized transport channels without significantly altering charge mobility.

Fig.~\ref{fig:wave_transition}(c,g) and Fig.~S1(a,b) illustrate the calculated spin conductivity and charge conductivity for both systems.
 A clear emergence of finite $\sigma^{z}_{xy}$ and $\sigma^{z}_{yx}$ components is observed under $u_{xx-yy}$ strain,
in agreement with the expected two-fold spin-splitting symmetry in the $d$-wave phase.
The overall magnitude of $\sigma^{z}_{ij}$ increases monotonically with strain,
confirming that mechanical symmetry reduction is an efficient way to induce nonrelativistic spin-polarized currents
in twisted bilayer altermagnets.

\begin{table*}[ht!]
	\centering
	\caption{
		Electric-field- and strain-induced responses of twisted bilayer altermagnets.
		ZSS$_{10}$ denotes the Zeeman-type spin splitting at an out-of-plane electric field
		$\mathcal{E}_z = 10$ MV/cm, reported as VBM / CBM energy differences.
		For biaxial strain ($-3\% \rightarrow +3\%$), the linear spin-splitting coefficient
		$\alpha^{(1)}$ is listed separately for the $K_c$ and $M_c$ points
		(hexagonal lattices) or the $M_c$ point (tetragonal lattices).
	}
	\label{tab:response_summary}
	\renewcommand{\arraystretch}{1.2}
	
	\resizebox{\textwidth}{!}{
		\begin{tabular}{lcccc}
			\toprule
			System &
			ZSS$_{10}$ (VBM / CBM) [meV] &
			$\alpha^{(1)}_{K_c}$ VBM / CBM &
			$\alpha^{(1)}_{M_c}$ VBM / CBM &
			Diagonal strain response \\
			\midrule
			
			tb-CoCl$_2$ (hex.) &
			38.5 / 37.9 &
			239 $\rightarrow$ 181 / 8.3 $\rightarrow$ 51.6 &
			619 $\rightarrow$ 68 / 11.4 $\rightarrow$ 6.4 &
			$i \rightarrow d$ \\
			
			tb-MnTe$_2$ &
			88.8 / 86.0 &
			203 $\rightarrow$ 146 / 442 $\rightarrow$ 21.5 &
			1046 $\rightarrow$ 209 / 877 $\rightarrow$ 383 &
			$i \rightarrow d$ \\
			
			tb-MnBr$_2$ &
			41.8 / 25.9 &
			244 $\rightarrow$ 198 / 416 $\rightarrow$ 223 &
			573 $\rightarrow$ 231 / 1303 $\rightarrow$ 954 &
			$i \rightarrow d$ \\
			
			tb-MnCl$_2$ &
			38.5 / 37.9 &
			156 $\rightarrow$ 106 / 309 $\rightarrow$ 171 &
			370 $\rightarrow$ 233 / 555 $\rightarrow$ 353 &
			$i \rightarrow d$ \\
			
			tb-MnI$_2$ &
			46.5 / 44.2 &
			648 $\rightarrow$ 531 / 310 $\rightarrow$ 249 &
			818 $\rightarrow$ 348 / 1363 $\rightarrow$ 1146 &
			$i \rightarrow d$ \\
			
			tb-MnSe$_2$ &
			9.9 / 39.8 &
			650 $\rightarrow$ 34 / 1027 $\rightarrow$ 96 &
			124 $\rightarrow$ 71 / 490 $\rightarrow$ 127 &
			$i \rightarrow d$ \\
			
			tb-VI$_2$ &
			45.6 / 45.6 &
			70 $\rightarrow$ 3.4 / 40 $\rightarrow$ 12.6 &
			23 $\rightarrow$ 29 / 653 $\rightarrow$ 276 &
			$i \rightarrow d$ \\
			
			tb-VCl$_2$ &
			37.8 / 37.5 &
			76 $\rightarrow$ 176 / 473 $\rightarrow$ 3.9 &
			139 $\rightarrow$ 120 / 499 $\rightarrow$ 183 &
			$i \rightarrow d$ \\
			
			tb-NiBr$_2$ &
			40.8 / 40.1 &
			465 $\rightarrow$ 412 / 97 $\rightarrow$ 150 &
			393 $\rightarrow$ 14 / 298 $\rightarrow$ 159 &
			$i \rightarrow d$ \\
			
			tb-NiF$_2$ &
			28.7 / 28.6 &
			46 $\rightarrow$ 45 / 41 $\rightarrow$ 5 &
			241 $\rightarrow$ 216 / 116 $\rightarrow$ 23 &
			$i \rightarrow d$ \\
			
			tb-CoCl$_2$ (tet.) &
			115.4 / 117.9 &
			-- &
			643 $\rightarrow$ 288 / 837 $\rightarrow$ 688 &
			$g \rightarrow d$ \\
			
			tb-FeS &
			38.4 / 14.0 &
			-- &
			27 $\rightarrow$ 389 / 46 $\rightarrow$ 233 &
			$d$ (no transition) \\
			
			tb-CoS &
			21.2 / 33.1 &
			-- &
			49 $\rightarrow$ 208 / 332 $\rightarrow$ 981 &
			$d$ (no transition) \\
			
			tb-RuSe &
			12.8 / 32.9 &
			-- &
			40 $\rightarrow$ 11 / 415 $\rightarrow$ 52 &
			$d$ (no transition) \\
			
			\bottomrule
		\end{tabular}
	}
\end{table*}

To quantify the charge-to-spin conversion efficiency, we also evaluated the spin-splitter angle (SSA),
defined as~\cite{karetta2025strain}
\begin{equation}
	\theta_{ij}^{(s)} = \tan^{-1} \left[ 2 \frac{2e}{\hbar} \frac{\sigma^{s}_{ij}}{\sigma^{c}_{ii}} \right],
\end{equation}
where $\sigma^{s}_{ij}$ and $\sigma^{c}_{ii}$ are the transverse spin and longitudinal charge conductivities, respectively.
For both tb-CoCl$_2$ and tb-MnTe$_2$, $\theta_{ij}^{(s)}$ is found to increase by approximately 7\textdegree and 18\textdegree, respectively, upon application of 1\% $u_{xx-yy}$ strain, starting from zero in the unstrained state.
The corresponding spin-splitter angle increases by several degrees under $1\%$ strain
[Fig.~\ref{fig:wave_transition}(d,h)], placing the effect well within experimentally measurable regimes. This enhancement reflects the emergence of the spin-splitter effect, 
analogous to the strain-induced spin current observed in altermagnetic CrSb~\cite{karetta2025strain}
and predicted for general collinear compensated magnets~\cite{smejkal2025tuning}.

Overall, these results demonstrate that the $u_{xx-yy}$ strain not only drives
the $g/i \rightarrow d$ altermagnetic transitions but also activates measurable spin-polarized transport responses,
without requiring spin–orbit coupling or magnetic canting. 
This establishes mechanical symmetry control as a viable mechanism for realizing 
nonrelativistic spin-splitter effects in two-dimensional altermagnets. Within the spin point group (SPG) framework, this diagonal strain removes the rotational operations that connect opposite spin sublattices, thereby lifting symmetry constraints on the spin-conductivity tensor and enabling finite $\sigma^z_{ij}$ components. In contrast, electric fields and biaxial strain primarily tune the magnitude of the existing nonrelativistic spin splitting without altering the underlying wave symmetry. A system-level overview of these symmetry--response correlations across the twisted bilayer family is summarized in Table~\ref{tab:response_summary}, which highlights how electric fields, biaxial strain, and diagonal strain collectively govern the evolution of NRSS and the activation of spin transport. Together, these findings establish mechanical symmetry control as a viable and general mechanism for realizing nonrelativistic spin-splitter effects in two-dimensional twisted altermagnets.

\section{Conclusion}

In summary, we have systematically investigated nonrelativistic spin splitting (NRSS) and its tunability in a broad class of twisted bilayer altermagnets, including CoCl$_2$, AX$_2$ (A = Mn, V, Ni; X = Cl, Br, I), FeS, CoS, MnTe$_2$, MnSe$_2$, and RuSe. 
Using first-principles calculations and spin-group symmetry analysis, we demonstrate that twisted stacking inherently breaks spin-compensating symmetries, giving rise to NRSS with distinct $d$-, $g$-, and $i$-wave altermagnetic patterns in momentum space. 
Out-of-plane electric fields induce Zeeman-type spin splitting up to $\sim$117 meV, whereas biaxial strain linearly tunes the linear spin-splitting coefficient $\alpha^{(1)}$ without altering symmetry. 
More crucially, diagonal strain ($u_{xx-yy}$) reduces rotational symmetries, driving $g/i \rightarrow d$ wave transitions that activate finite spin conductivity and enhance spin-splitter angles by $\sim$18\textdegree. 
These effects, observed across different lattice types, confirm that NRSS and spin transport can be efficiently controlled without spin–orbit coupling. 
Our findings establish twisted bilayer altermagnets as a general platform for exchange-driven, symmetry-tunable spin transport, providing a pathway toward spin–orbit-free spintronic and straintronic devices based on light-element 2D materials.

\section{Computational Methods}

In the nonrelativistic regime, charge and spin conductivities are expressed through spin-resolved channels corresponding to spin-up ($s_k$) and spin-down ($-s_k$) polarizations along the quantization axis $k$. 
The spin-polarized conductivity tensor $\sigma_{ij}^{s_k}$ relates to the current density $J_i^{s_k}$ as $J_i^{s_k} = \sigma_{ij}^{s_k}\mathcal{E}_j$, where $\mathcal{E}_j$ is the electric field along direction $j$. 
Within the semiclassical Boltzmann transport framework~\cite{callaway2013quantum}, the spin-resolved conductivity can be written as
\begin{equation}
	\sigma^{s_k}_{ij}(E_F) = -\frac{e^2 \tau}{8\pi^3 \hbar^2} 
	\sum_n \int 
	\frac{\partial E^{s_k}_{n\mathbf{k}}}{\partial k_i}
	\frac{\partial E^{s_k}_{n\mathbf{k}}}{\partial k_j}
	\frac{\partial f^0}{\partial E^{s_k}_{n\mathbf{k}}}
	\, d^3k,
\end{equation}
where $\tau$ denotes the relaxation time and $f^0$ is the Fermi–Dirac distribution. 
The total charge and spin conductivities are then
\begin{equation}
	\begin{aligned}
		\sigma_{ij} &= \sum_{k}\big(\sigma^{s_k}_{ij} + \sigma^{-s_k}_{ij}\big), \\
		\sigma^k_{ij} &= \frac{\hbar}{2e}\big(\sigma^{s_k}_{ij} - \sigma^{-s_k}_{ij}\big).
	\end{aligned}
\end{equation}
Under time-reversal operation $\mathcal{T}$, $\sigma^{s_k}_{ij}$ transforms as $\mathcal{T}\sigma^{s_k}_{ij}\rightarrow \sigma^{-s_k}_{ij}$, rendering $\sigma_{ij}$ even and $\sigma^k_{ij}$ odd with respect to $\mathcal{T}$. 
In the absence of spin–orbit interaction (SOI), spin canting is suppressed, leading to vanishing $\sigma^{x/y}_{ij}$ components, while $\sigma^{z}_{ij}$ remains finite. 
This can be verified through the spin-only symmetry operation $[C_z(\pi)||E]$, which inverts the in-plane spin current components, ensuring that $\sigma^{x/y}_{ij}=0$. 

All calculations were carried out within density functional theory (DFT) using the \textsc{VASP} package~\cite{kresse1996vasp1,kresse1996vasp2}. 
The ion–electron interaction was described by the projector augmented-wave (PAW) method~\cite{blochl1994projector}. 
Exchange–correlation effects were treated with the Perdew–Burke–Ernzerhof (PBE) functional within the generalized gradient approximation (GGA)~\cite{perdew1996generalized}. 
A plane-wave kinetic energy cutoff of 520 eV was used, and the Brillouin zone was sampled using a $\Gamma$-centered Monkhorst–Pack $k$-mesh with a spacing of 0.02 Å$^{-1}$. 
All atomic positions and lattice parameters were optimized until the residual force on each atom was less than 0.001 eV/Å, and total energy convergence was better than $10^{-6}$ eV. 
A vacuum region exceeding 15 Å was introduced to prevent interlayer coupling between periodic images.

To account for long-range van der Waals (vdW) interactions, the DFT-D3 correction of Grimme was employed~\cite{grimme2010consistent}. 
The GGA+$U$ approach~\cite{dudarev1998electron} was used to incorporate on-site Coulomb interaction in the transition-metal $d$ orbitals, with $J=0$. 
Following previous studies on monolayer and bilayer transition-metal halides and chalcogenides, the following $U$ values were adopted:
$U_{\mathrm{Mn}} = 4.0$ eV~\cite{gong2017discovery}, 
$U_{\mathrm{Co}} = 3.5$ eV~\cite{kong2022strain}, 
$U_{\mathrm{V}} = 3.0$ eV~\cite{xu2020electronic},
$U_{\mathrm{Ni}} = 6.0$ eV~\cite{huang2020nihalides}, 
$U_{\mathrm{Fe}} = 5.0$ eV~\cite{zhang2019fehalides}, and 
$U_{\mathrm{Ru}} = 2.5$ eV~\cite{belashchenko2025giant}. 
All qualitative results remain unchanged under reasonable variations of $U$.

A uniform out-of-plane electric field ($\mathcal{E}_z$) was introduced using the dipole correction method of Neugebauer and Scheffler~\cite{neugebauer1992adsorbate}, which places a compensating dipole sheet in the vacuum region to eliminate artificial periodic image interactions. 

Maximally localized Wannier functions (MLWFs) were constructed using the \textsc{Wannier90} code~\cite{mostofi2008wannier90} from VASP wavefunctions, using transition-metal $d$ and ligand $p$ orbitals as projectors. 
The resulting tight-binding Hamiltonian was employed within the \textsc{WannierBerri} framework~\cite{tsirkin2021wannierberri} to compute the nonrelativistic spin and charge conductivities. 
The spin conductivity tensor in the Kubo formalism is given as
\begin{equation}
	\begin{aligned}
		\sigma^{k}_{ij} = 
		- \frac{e\hbar}{2\Gamma V}
		\int \frac{d^3\mathbf{k}}{(2\pi)^3}
		\sum_n
		\langle n\mathbf{k} | \hat{J}^k_i | n\mathbf{k} \rangle
		\langle n\mathbf{k} | \hat{v}_j | n\mathbf{k} \rangle
		\delta(E_{n\mathbf{k}} - E_F),
	\end{aligned}
\end{equation}
where $\hat{J}^k_i = \frac{1}{2}\{\hat{v}_i, \hat{s}_k\}$ denotes the spin-current operator and $\Gamma=\hbar/(2\tau)$ is the scattering rate. 
Replacing $\hat{J}^k_i$ with $\hat{v}_i$ yields the charge conductivity tensor. 
Integrations were performed over a dense $400\times400$ $k$-mesh for 2D systems to achieve convergence of $\sigma^{z}_{ij}$.

All symmetry analyses were performed using \textsc{FINDSYM}~\cite{stokes2005findsym}, the Bilbao Crystallographic Server~\cite{aroyo2006bilbao,aroyo2011crystallography}, \textsc{FINDSPINGROUP}~\cite{chen2024_spinsg_prx}, and the \textsc{TensorSymmetry} package~\cite{xiao2025tensorsymmetry}. 
Additional visualization and pre/post-processing were carried out using \textsc{VASPKIT}~\cite{wang2021vaspkit}, \textsc{PyProcar}~\cite{herath2020pyprocar}, \textsc{SEEK-PATH}~\cite{hinuma2017seekpath}, \textsc{MAGNDATA}~\cite{gallego2016magndata}, and the Materials Project database~\cite{jain2013materials}. 
These resources were used to determine the spin point groups, verify transformation rules for $\sigma^{z}_{ij}$, and ensure consistency with the analytical symmetry models.

\section*{Supporting Information}

The Supporting Information includes (i) charge conductivity $\sigma^{z}_{xx}$ and $\sigma^{z}_{yy}$ as a function of energy for twisted bilayer CoCl$_2$ and MnTe$_2$ under unstrained and strained conditions; (ii) the linear nonrelativistic spin-splitting coefficient $\alpha^{(1)}$ (meV\,\AA) as a function of biaxial strain for representative twisted bilayers; and (iii) electric-field-induced spin splitting $\Delta$ at the VBM and CBM as a function of out-of-plane electric field $\mathcal{E}_z$ for representative hexagonal and tetragonal systems.

\section*{Acknowledgments}  
S.P. acknowledges PMRF, India, for the Research fellowship [Grant No. 1403227]. S. B. acknowledges financial support from SERB under a core research grant (grant no. CRG/2019/000647) to set up his High Performance Computing (HPC) facility “Veena” at IIT Delhi for computational resources.

\bibliography{ref}

\end{document}